\documentclass[fleqn,usenatbib,useAMS]{mnras}
\usepackage{amsfonts,amsmath,units,wasysym,epsfig,graphicx,verbatim,color,subfigure,graphicx,bm,mathrsfs,lipsum,hyperref}

\usepackage[normalem]{ulem}  
\definecolor{dgreen}{RGB}{0, 204, 0}

\newcommand{\IUCAA}{Inter-University Centre for Astronomy and Astrophysics, Post Bag 4, Ganeshkhind, Pune 411 007, India}
\newcommand{\LIEGE}{Space  sciences,  Technologies  and  Astrophysics  Research  (STAR)  Institute,  Université de Liège, Bât. B5a, 4000 Liège, Belgium} 
\newcommand{\PM}{Polba Mahavidyalaya, Hooghly, West Bengal 712148, India}  
\newcommand{\WSU}{Department of Physics \& Astronomy, Washington State University, 1245 Webster, Pullman, WA 99164-2814, U.S.A}
\newcommand{\AUT}{Aristotle University of Thessaloniki, Department of Physics, University Campus Laboratory of Astronomy, 54124 Thessaloniki, Greece}

\begin{document}
\label{firstpage}
\pagerange{\pageref{firstpage}--\pageref{lastpage}}

\title{GW190814: On the properties of the secondary component of the binary}

\author[Bhaskar Biswas et al.]{Bhaskar Biswas,$^{\rm 1}$ Rana Nandi,$^{\rm 2}$ Prasanta Char,$^{\rm 3}$ Sukanta Bose,$^{\rm 1, 4}$ and Nikolaos Stergioulas$^{\rm 5}$\\
$^{\rm 1}$\IUCAA \\
$^{\rm 2}$\PM \\ 
$^{\rm 3}$\LIEGE \\
$^{\rm 4}$\WSU \\
$^{\rm 5}$\AUT}

\maketitle
\begin{abstract}
We show that the odds of the mass-gap (secondary) object in GW190814 being a neutron star (NS) improve if one allows for a stiff high-density equation of state (EoS) or a large spin.
Since its mass is $\in (2.50,2.67) M_{\odot}$, establishing its true nature will make it either the heaviest NS or the lightest black hole (BH), and can have far-reaching implications on NS EoS and compact object formation channels. When limiting oneself to the NS hypothesis, we deduce the secondary's properties by
using a Bayesian framework with a hybrid EoS formulation that  employs a parabolic expansion-based nuclear empirical parameterization around the nuclear saturation density augmented by a generic 3-segment piecewise polytrope (PP) model at higher densities and combining a variety of  astrophysical observations. For the slow-rotation scenario, GW190814 implies a very stiff EoS and a stringent constraint on the EoS specially in the high-density region.  On the other hand assuming the secondary object is a rapidly rotating NS, we constrain its rotational frequency to be $f=1170^{+389}_{-495}$ Hz, within a $90\%$ confidence interval (CI). In this scenario, the secondary object in GW190814 would qualify as the fastest rotating NS ever observed.  However, for this scenario to be viable, rotational instabilities would have to be suppressed both during formation and the subsequent evolution until merger, otherwise the secondary of GW190814 is more likely to be a BH. 
\end{abstract}


\section{Introduction}
Recently, the LIGO~\citep{advanced-ligo} and Virgo~\citep{advanced-virgo} scientific collaborations (LVC) reported 
the detection of one of the most enigmatic gravitational wave (GW) mergers till date~\citep{Abbott:2020khf}. 
This event, named GW190814, has been associated with a compact object binary of mass-ratio, $q = 0.112^{+0.008}_{-0.009}$, and primary and secondary masses $m_1 = 23.2^{+1.1}_{-1.0} M_\odot$ and $m_2 = 2.59^{+0.08}_{-0.09}$, respectively. Since, an electromagnetic (EM) counterpart has not been found for this particular event and the tidal deformability has not been measurable from the GW signal, the secondary component might well be the lightest BH ever found.
However, EM emissions are expected to be observed for only a fraction of NS binaries, and tidal deformabilities are known to be small for massive NSs, hence the secondary in this case cannot be ruled out as a NS.
In the latter scenario, it would become the
heaviest NS observed in a binary system, given its  well-constrained mass. Either hypothesis deserves a deep study owing to its far-reaching implications on the formation channels of such objects and the nature of the densest form of matter in the universe.

Discoveries of massive pulsars in past decades have severely constrained the EoS of supranuclear matter inside their cores~\citep{Demorest:2010bx,Antoniadis:2013pzd,Fonseca:2016tux,Arzoumanian:2017puf,Cromartie:2019kug}. These observations provided a very strong lower bound of $\sim 2 M_\odot$ on the maximum mass 
of nonrotating NSs that all the competing EoS models from nuclear physics must satisfy. Furthermore, GW170817~\citep{TheLIGOScientific:2017qsa} has prompted several studies predicting an upper bound of $\sim 2.2-2.3 M_\odot$ on $M_{\rm max}$ of nonrotating NSs, based on the mass ejecta, kilonova signal and absence of a prompt collapse \citep{Shibata:2017xdx, Margalit:2017dij, Ruiz:2017due, Rezzolla:2017aly, Shibata:2019ctb, PhysRevD.101.063029}.  While the simultaneous mass-radius measurements of PSR J0030+0451 by NICER collaboration \citep{Riley:2019yda,Miller:2019cac} indicate a tilt towards slightly stiffer EoS \citep{Raaijmakers:2019dks, Landry_2020PhRvD.101l3007L,Biswas_arXiv_2008.01582B}, 
the distribution of $m_2$ would require even higher $M_{\rm max}$. Possible formation channels of GW190814-type binaries have also been studied in some recent works \citep{Zevin:2020gma,Safarzadeh_2020ApJ...899L..15S,Kinugawa_2020arXiv200713343K}. While there is a general consensus that the fallback of a significant amount of bound supernova ejecta on the secondary compact remnant leads to its formation in the lower mass-gap region, the nature of its state at the time of the merger being a BH or a NS remains unclear. Nevertheless, GW190814 has motivated experts to reevaluate the knowledge of dense matter and stellar structure to determine the possible scenarios in which one can construct such configurations of NSs while satisfying  relevant constraints \citep{Most_2020MNRAS.499L..82M, Zhang_2020ApJ...902...38Z, Fattoyev_2020PhRvC.102f5805F, Tsokaros_2020ApJ...905...48T, Tews_2021ApJ...908L...1T, Lim_2020arXiv200706526L, Dexheimer_arXiv_2007.08493D, Sedrakian_2020PhRvD.102d1301S, Godzieba_arxiv_2020.10999, Huang_2020ApJ...904...39H, Demircik:2020jkc,LI2020135812}. Most of these works suggest rapid uniform rotation with or without exotic matter, such as hyperons or quark matter, exploiting the caveat that the spin of $m_2$ is unconstrained. Other  possibilities such as $m_2$ being a primordial BH~\citep{Vattis:2020iuz,Jedamzik_2021PhRvL.126e1302J,Clesse_arXiv_2007.06481C}, an anisotropic object~\citep{Roupas_2021Ap&SS.366....9R} [see also~\citep{Biswas:2019gkw} for a detailed study on anisotropic object] or a NS in scalar-tensor gravity~\citep{Rosca-Mead_2020Symm...12.1384R} have also been considered. 

In this article, we investigate the possibility of the GW190814's secondary being a NS within a hybrid nuclear+PP EoS parameterization~\citep{Biswas_arXiv_2008.01582B}, and study its related properties under  assumptions of it being both slowly and rapidly rotating. We also constrain its spin using a universal relation developed by \citet{Breu:2016ufb}.

\section{A brief review of our previous work}
\label{previuos-work}
In a previous work~\citep{Biswas_arXiv_2008.01582B}, we have employed Bayesian statistics to constrain the EoS of NS combining multiple astrophysical observations. We have formulated a hybrid nuclear+PP EoS model which uses a parabolic expansion based nuclear empirical parameterization around the nuclear saturation density ($\rho_0$) and a 3-segment PP parameterization at higher densities. Within the parabolic expansion, the energy per nucleon $e (\rho, \delta)$ of asymmetric nuclear matter can be expressed as:
\begin{equation}
    e(\rho,\delta) \approx  e_0(\rho) +  e_{\rm sym}(\rho)\delta^2,
\end{equation}
where $e_0(\rho)$ is the energy of the symmetric nuclear matter which holds equal number of neutrons and protons. The $e_{\rm sym}(\rho)$ is known as the symmetry energy which characterizes the strength of asymmetry in neutron to proton ratio, and $\delta=(\rho_n-\rho_p)/\rho$ is known as symmetry parameter. $e_0(\rho)$ and $e_{\rm sym}(\rho)$ can be further expanded in a Taylor series around $\rho_0$: 
\begin{eqnarray}
 e_0(\rho) &=&  e_0(\rho_0) + \frac{ K_0}{2}\chi^2 \label{eq:e0} +\,...,\\
e_{\rm sym}(\rho) &=&  e_{\rm sym}(\rho_0) + L\chi + \frac{ K_{\rm sym}}{2}\chi^2 
 + ..., \label{eq:esym}
\end{eqnarray}
where $\chi \equiv (\rho-\rho_0)/3\rho_0$.

At higher densities the EoS of nuclear matter is completely unknown to us. This is the reason we choose a generic 3-segment PP parameterization after $1.25 \rho_0$. This particular transition density is motivated by  Bayesian evidence calculation which is detailed in~\cite{Biswas_arXiv_2008.01582B}. Then, we construct the posterior of the EoS parameters using Bayesian statistics based on this hybrid nuclear+PP model by combining astrophysical data from the radio observation of PSR J0740+6620, GW170817, GW190425, and NICER observations: 
\begin{equation}
    P(\theta | {d}) = \frac{P ({d} | \theta) \times P(\theta)}{P(d)}\, = \frac{\Pi_i P ({d_i} | \theta) \times P(\theta)}{P(d)}\,,
    \label{bayes theorem}
\end{equation}

where $\theta$ is the set of EoS parameters in the model, $d = (d_{\rm GW}, d_{\rm X-ray}, d_{\rm Radio})$ is the set of data from the three different 
types of observations that are used to construct the likelihood. The mathematical expressions to compute each of the individual likelihoods are given in Eq. 5, 6, and 7 of ~\cite{Biswas_arXiv_2008.01582B} respectively.

In this present paper, we make use the methodology built in our previous work and investigate the properties of the secondary object of GW190814 under a variety of assumption.

\section{Lightest BH or heaviest NS?}
 The mass of the secondary object in GW190814 measured by the LVC falls into the so called ``mass gap" region~\citep{Bailyn_1998,_zel_2010} and, therefore, demands a careful inspection of its properties before it can be ruled out as 
a BH or NS.

\begin{figure}
    \centering
    \includegraphics[width=0.45\textwidth]{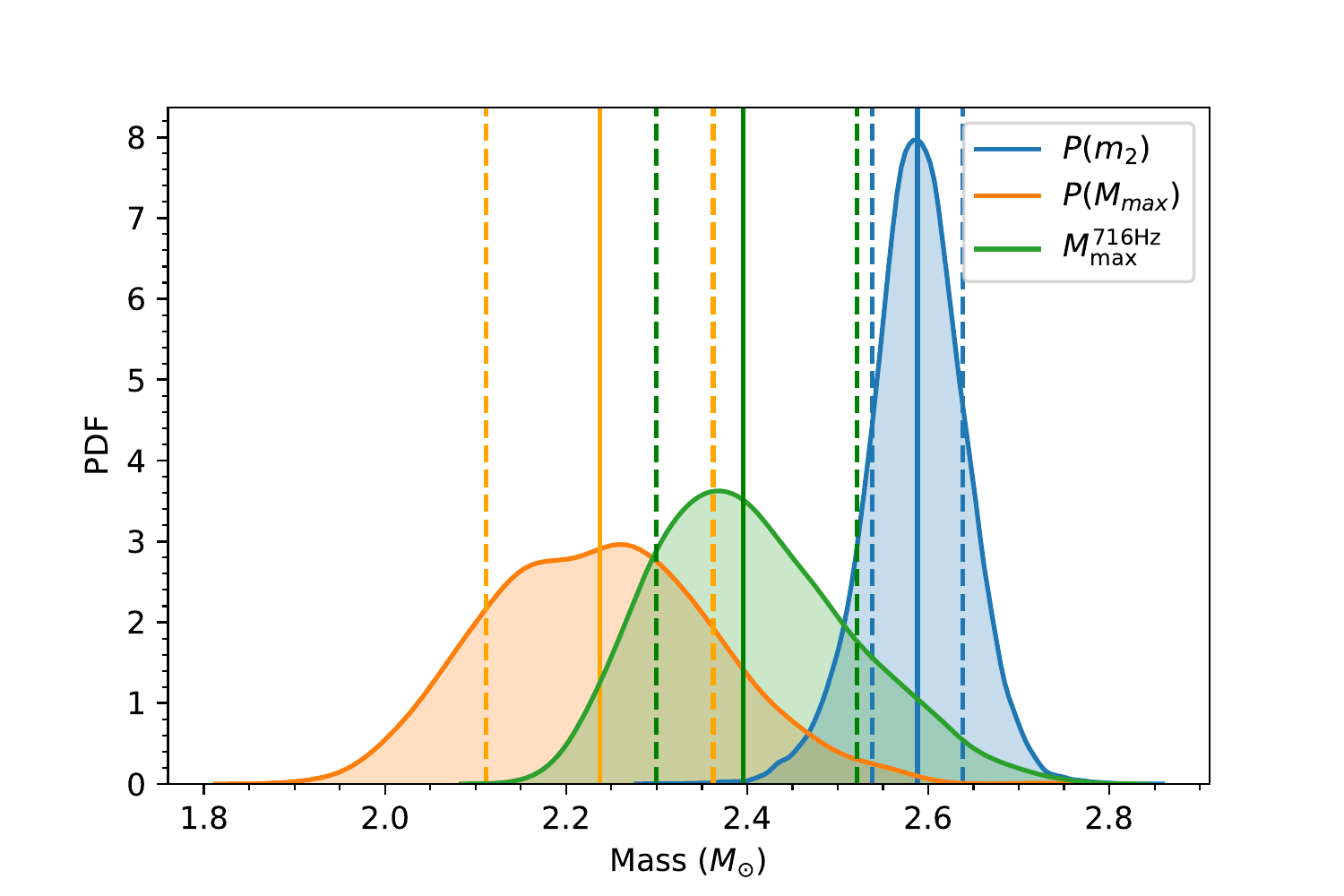}
    \caption{The probability distribution of $M_{\rm max}$ of NSs, obtained from~\citet{Biswas_arXiv_2008.01582B}, is
    shown in orange. The distribution shown in green is obtained with the same EoS samples as for the orange one, but considering uniform NS rotation  at 716 Hz. These two distributions are
    compared with the probability distribution of the secondary's mass $m_2$ (in blue) deduced from the GW190814 posterior samples in ~\citet{Abbott:2020khf}. }
    \label{fig:BHNS-prob}
\end{figure}

A non-informative measurement of the tidal deformability or the spin of the secondary,
or the absence of an EM counterpart associated with this event, have made it difficult to make a robust statement about the nature of this object. 
We begin by examining if the GW mass measurement along with hybrid nuclear+PP model alone 
can rule it out as a NS.
In Fig.~\ref{fig:BHNS-prob} the posterior distribution of secondary mass $m_2$ is plotted, in blue, by using publicly available LVC posterior samples~\footnote{LVK collaboration,~\href{https://dcc.ligo.org/LIGO-P2000183/public}{https://dcc.ligo.org/LIGO-P2000183/public}}. In orange, the posterior distribution of $M_{\rm max}$ is overlaid from hybrid nuclear+PP model analysis by~\citet{Biswas_arXiv_2008.01582B} using PSR J0740+6620~\citep{Cromartie:2019kug}, combined GW170817~\footnote{LVK collaboration,~\href{https://dcc.ligo.org/LIGO-P1800115/public}{https://dcc.ligo.org/LIGO-P1800115/public}} and GW190425~\footnote{LVK collaboration,~\href{https://dcc.ligo.org/LIGO-P2000026/public}{https://dcc.ligo.org/LIGO-P2000026/public}}, and NICER~\footnote{PSR J0030+0451 mass-radius samples released by ~\citet{Miller:2019cac},~\href{https://zenodo.org/record/3473466\#.XrOt1nWlxBc}{https://zenodo.org/record/3473466\#.XrOt1nWlxBc}} data.   

Given these two distributions 
-- both for nonrotating stars -- 
we calculate the probability of $m_2$ being greater than $ M_{\rm max}$, i.e., $P(m_2 > M_{\rm max})$ = $P(m_2 - M_{\rm max})$. This probability can be easily obtained by calculating the convolution of the $m_2$ and $- M_{\rm max}$ probability distributions, which yields $P(m_2 > M_{max}) = 0.99$. Therefore, the mass measurement implies that the probability that the secondary object in GW190814 is a NS is $\sim 1\%$. However, this type of analysis is highly sensitive to the choice of EoS parameterization as well as on the implementation of the  maximum-mass constraint obtained from the heaviest pulsar observations. The LVC analysis~\citep{Abbott:2020khf} which is based on the spectral EoS parameterization~\citep{Lindblom:2010bb}, obtained~$\sim 3\%$ probability for the secondary to be a NS using GW170817-informed EoS samples from~\citet{Abbott:2018exr}. The addition of NICER data might increase this probability. ~\citet{Essick_2020ApJ...904...80E} added NICER data in their analysis of GW observations based on a nonparametric EoS and also examined the impact of different assumptions about the compact object mass distribution. The $P(m_2 > M_{max})$ probabilities technically depend on the mass prior assumed for the secondary, but ~\citet{Essick-2020arXiv} showed that, regardless of assumed population model, there is a less than $\sim 6\%$ probability for the GW190814 secondary to be a NS. In the discovery paper, LVC also reported an EoS-independent result using the pulsar mass distribution, following ~\cite{Farr2020}, which suggests that there is less than $\sim 29\%$ probability that the secondary is a NS. Despite the differences inherent to these studies, they all
suggest that there is a small but finite probability of the secondary object in GW190814 to be a NS. It is also important to note that they
all assumed the NS to be either nonrotating or slowly rotating ($\chi < 0.05$).

Another possibility is that the secondary object is a rapidly rotating NS~\citep{Most_2020MNRAS.499L..82M,Tsokaros_2020ApJ...905...48T}. It is known that uniform rotation  can increase the maximum mass of a NS by $\sim20 \%$~\citep{1987ApJ...314..594F,Cook-a,Cook-b}. Therefore, rapid rotation may 
improve the chances that the GW190814 data are consistent with a NS.

From pulsar observations, we know that NSs with spin frequencies as high as $\nu^{\rm obs}_{\rm max} = 716$ Hz exist in nature~\citep{Hessels:2006ze}. Using this value for the spin frequency and the EoS samples of~\citet{Biswas_arXiv_2008.01582B} we can deduce the maximum improvement in probability that the GW190814 secondary is a NS.
We used this information in the {\tt RNS} code~\citep{Stergioulas:1994ea} and obtained a corresponding distribution of maximum mass denoted as $M_{\rm max}^{716 \rm Hz}$. The superscript ``716 Hz" emphasizes that all configurations here are computed at that fixed spin frequency. 
In Fig.~\ref{fig:BHNS-prob}, the distribution of $M_{\rm max}^{716 \rm Hz}$ is shown in green. From the overlap of this distribution with $P (m_2)$, we find there is $\sim 8 \%$ probability that $m_2$ is a rapidly rotating NS. 

Alternatively, if the GW190814's secondary were indeed a NS, then the LVC mass measurement sets a lower limit on the maximum NS mass for any spin at least up to $\nu^{\rm obs}_{\rm max}$.

We next relax 
this constraint by considering all theoretically allowed values of the spin frequency,
which for some masses and EoSs may exceed the maximum observed value.
In the next two sections,  
we investigate the properties of NSs -- for various rotational frequencies -- 
using a Bayesian approach based on hybrid nuclear+PP EoS parameterization.

\section{Properties assuming a slowly rotating NS}
\begin{figure*}
    \centering
    \includegraphics[width=0.9\textwidth]{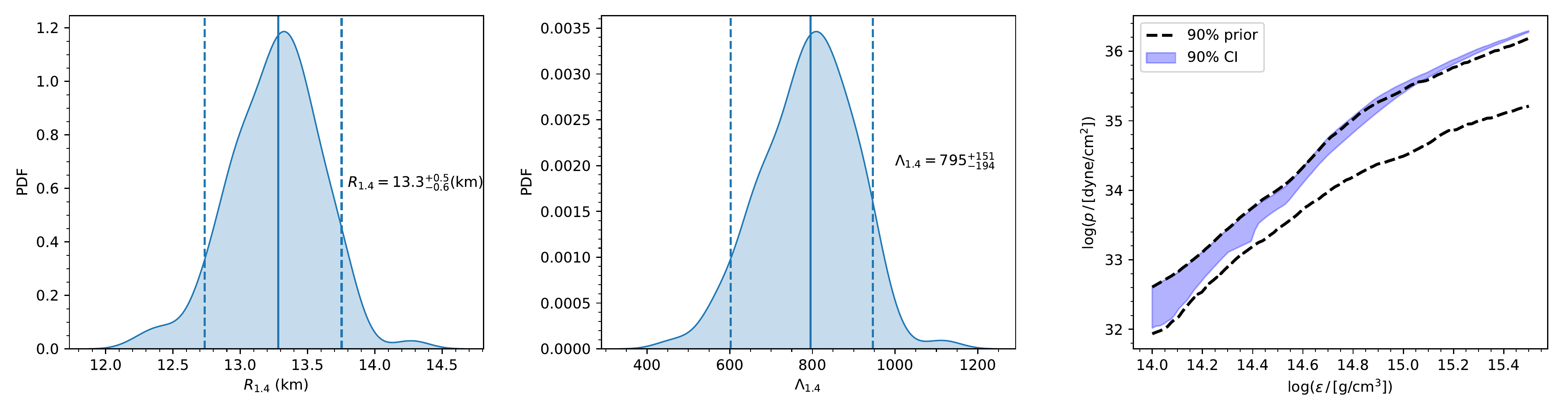}
    \caption{Posterior distributions of $R_{1.4}$ (left panel) and $\Lambda_{1.4}$ (middle panel), as well as the pressure as a function of energy density (right panel) are plotted  assuming that the secondary companion of GW190814 is a nonrotating NS. Median and $90\%$ CI are shown by solid and dashed lines, respectively.}
    \label{fig:nonrotating-prop}
\end{figure*}

 For slowly rotating NS, a Bayesian methodology was already developed in~\citet{Biswas_arXiv_2008.01582B} (also briefly described in Sec.~\ref{previuos-work} ) by combining multiple observations based on hybrid nuclear+PP EoS parameterization. In this paper, instead of marginalizing over the mass of PSR J0740+6620 taking into account of its measurement uncertainties (as described in~\citet{Biswas_arXiv_2008.01582B}), we consider the $m_2$ distribution of GW190814 as the heaviest pulsar mass measurement. We use Gaussian kernel-density to approximate the posterior distribution of $m_2$.
The resulting posteriors of radius ($R_{1.4}$) and tidal deformability ($\Lambda_{1.4}$) obtained from this analysis are plotted in Fig.~\ref{fig:nonrotating-prop}. 
We find that $R_{1.4}=13.3^{+0.5}_{-0.6} $km and $\Lambda_{1.4}=795^{+151}_{-194}$, at $90 \%$ CI, which are in good agreement with previous studies~\citep{Abbott:2020khf,Essick_2020ApJ...904...80E,Tews_2021ApJ...908L...1T}. 

The addition of GW190814 makes the EoS stiffer, especially in the high density region since now a very small subspace of the EoS family can support a $\sim 2.6M_{\odot}$ NS. In the right panel of Fig.~\ref{fig:nonrotating-prop}, the $90 \%$ CI of posterior of the pressure inside the NS is plotted as a function of energy density in shaded blue colour and the corresponding $90 \%$ CI of prior is shown by the black dotted lines. This plot clearly shows that the addition of GW190814 places a very tight constraint on the high-density part of the EoS. 

\section{Properties assuming a rapidly rotating NS}
\label{rapid-rotation}
In this article, for the first time, we develop a Bayesian formalism to constrain the EoS of NS that allows for rapid rotation. 
We use a universal relation found by~\citet{Breu:2016ufb} which relates the maximum mass of a uniformly rotating star ($M_{\rm rmax}^{\rm rot}$) with the maximum mass of a nonrotating star ($M_{\rm max}^{\rm TOV}$) for the same EoS,
    \begin{equation}
        M_{\rm rmax}^{\rm rot} = M_{\rm max}^{\rm TOV} \left[1+a_1\left(\frac{\chi}{\chi_{\rm kep}}\right)^2 +a_2\left(\frac{\chi}{\chi_{\rm kep}}\right)^4\right]\, ,
        \label{breu and rezolla:universal relation}
    \end{equation}
where $a_1=0.132$ and $a_2=0.071$. $\chi$ is the dimensionless spin magnitude of a uniformly rotating star and $\chi_{\rm kep}$ is the maximum allowed dimensionless spin magnitude at the mass-shedding limit. Given a $\chi/\chi_{\rm kep}$ value, we calculate $M_{\rm rmax}^{\rm rot}$ using this universal relation. Its use makes our computation much faster but can cause up to $\sim 2\%$ deviation from the exact result, as noted by~\citet{Breu:2016ufb}. We assume that the error is constant throughout the parameter space; we took it to be distributed uniformly in $[-2\%,2\%]$ and marginalized over it to get an unbiased estimate of the properties of the object. 

We combine data from PSR J0740+6620, two other binary neutron stars, namely GW170817 and GW190425, as well as NICER data assuming nonrotating NS following~\citet{Biswas_arXiv_2008.01582B}. 
Then, the $m_2$ distribution of GW190814 is used for the maximum-mass threshold of a uniformly rotating star, i.e., $M_{\rm rmax}^{\rm rot}$.  We use a nested sampler algorithm implemented in {\tt Pymultinest}~\citep{Buchner:2014nha} to simultaneously sample the EoS parameters and $\chi/\chi_{\rm kep}$. These posterior samples are then used in the {\tt RNS} code~\citep{Stergioulas:1994ea} to calculate several properties of the secondary object associated with GW190814. 
\begin{figure*}
    \centering
    \includegraphics[width=0.9\textwidth]{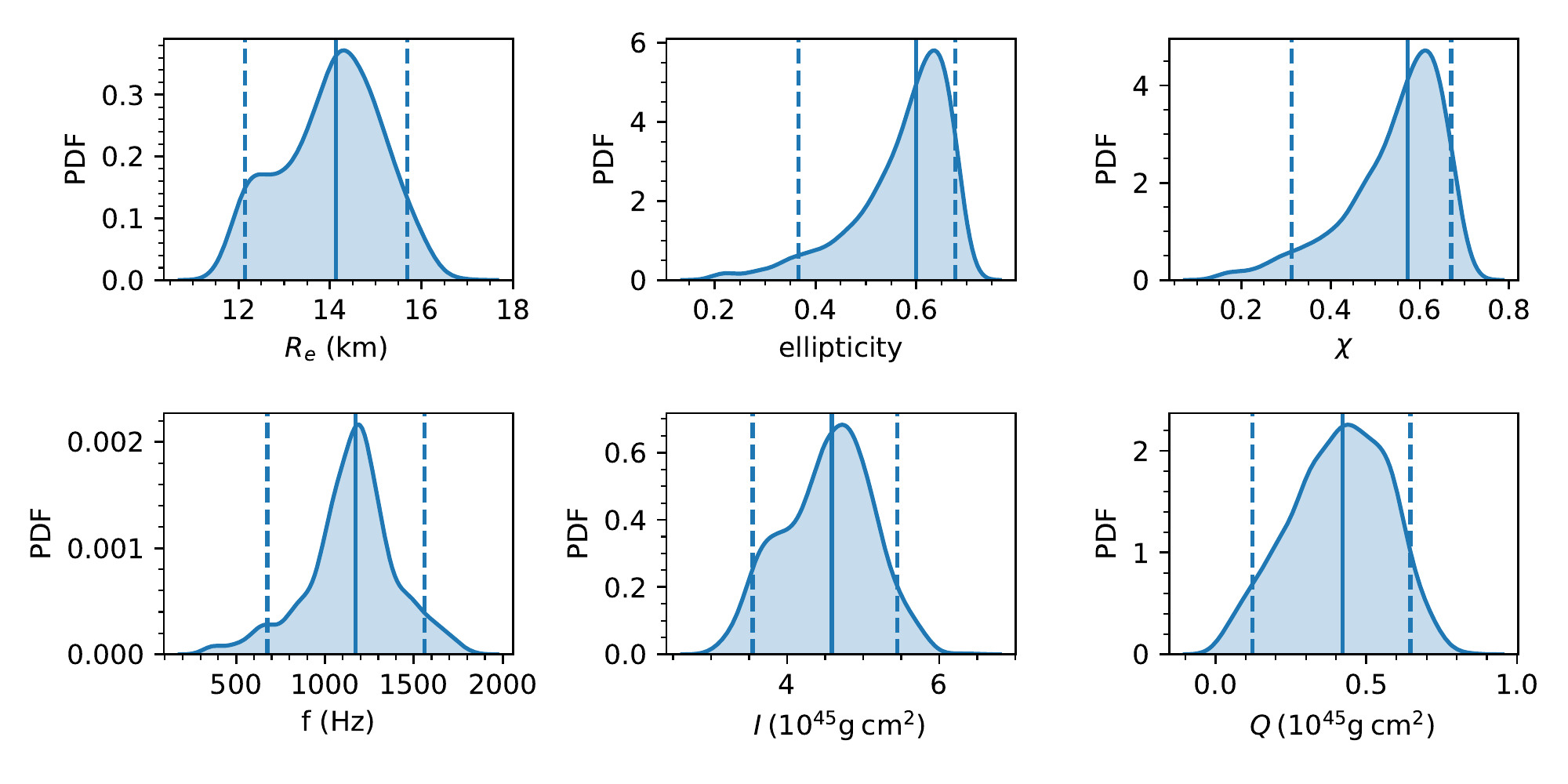}
    \caption{Posterior distribution of various properties of the secondary companion of GW190814 are shown assuming a rapidly rotating NS:  Equatorial radius $R_e$ (upper left), ellipticity $e$ (upper middle), dimensionless spin magnitude $\chi$ (upper right), rotational frequency $f$ in Hz (lower left),  moment of Inertia $I$ (lower middle) and  quadrupole moment $Q$ (lower right). Median and $90\%$ CI are shown by solid and dashed lines, respectively.}
    \label{fig:rotating-prop}
\end{figure*}

In the upper left and middle panel of Fig.~\ref{fig:rotating-prop} posterior distributions of equatorial radius ($R_e$) and ellipticiy ($e$) are plotted, respectively. Within the $90\%$ CI we find $R_e =14.1^{+1.5}_{-2.0} $km and $e = 0.60^{+0.07}_{-0.23}$. Such high values of equatorial radius and ellipticity imply a considerable deviation from a spherically symmetric static configuration.
From the distribution of $\chi$ shown in the upper left panel of Fig.~\ref{fig:rotating-prop} we find its value to be $\chi = 0.57^{+0.09}_{-0.26}$. \citet{Most_2020MNRAS.499L..82M} have also obtained a similar bound on $\chi$ with simpler arguments. In this paper, we provide a distribution for $\chi$ employing a Bayesian framework as well as place a more robust bound on this parameter.
In the lower left panel of Fig.~\ref{fig:rotating-prop}, the posterior distribution of rotational frequency is plotted in Hz. We find its value to be $f=1170^{+389}_{-495}$ Hz. As noted above, till date PSR J1748–2446a~\citep{Hessels:2006ze} is known as the fastest rotating pulsar, with a rotational frequency of 716 Hz. {\em Therefore, if the secondary of GW190814 is indeed a rapidly rotating NS, it would definitely be the fastest rotating NS observed so far.} In the lower-middle and right panels, the posterior distributions of the moment of inertia and quadrupole moment of the secondary are shown, respectively.

\subsection{Maximum spin frequencies and rotational instabilities}
\begin{figure*}
    \centering
    \includegraphics[width=\textwidth]{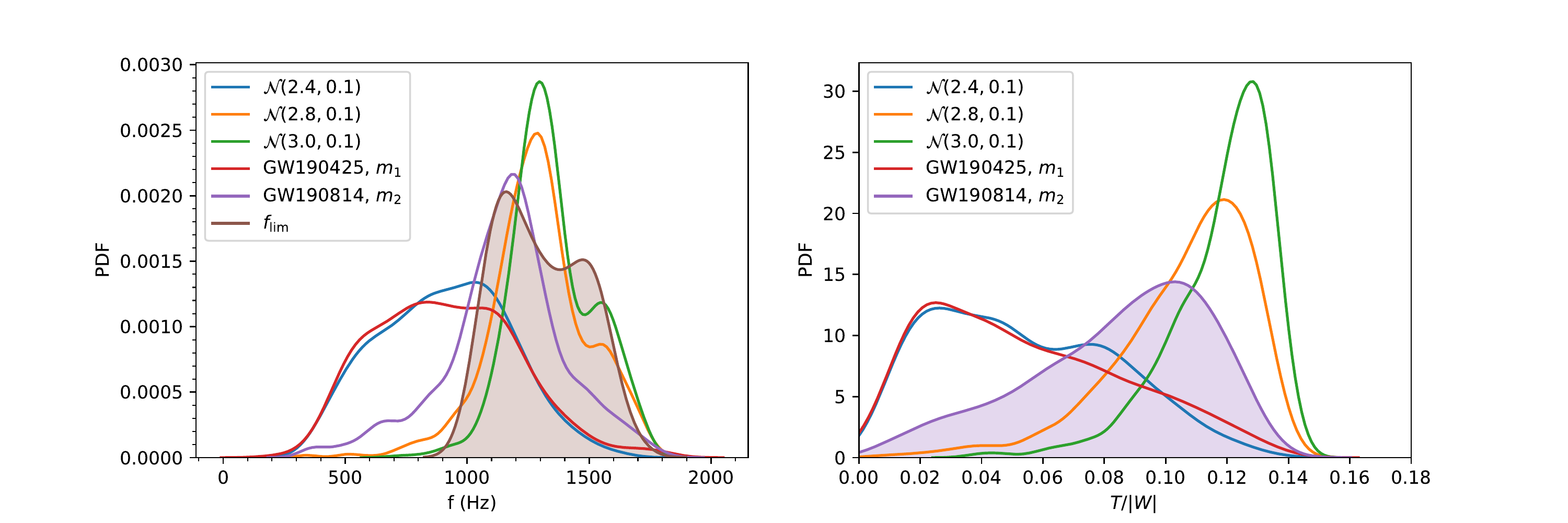}
    \caption{In the left panel, the probability distribution of $f_{\rm lim}$ is shown in brown shade. The distribution of $f_{\rm lim}$ is plotted considering three simulated rapidly rotating NS whose mass measurements are Gaussian distributed with median 2.4 $M_{\odot}$, 2.8 $M_{\odot}$ and 3.0 $M_{\odot}$, respectively and each having a measurement uncertainty of $.1 M_{\odot}$. The same has been overlaid using the secondary component of the GW190814 and the primary of the GW190425 events. In the right panel, the  corresponding ratio of rotational to gravitational potential energy $T/|W|$ is shown. 
    }
    
    \label{fig:rot-NS-prob}
\end{figure*}

EoS constraints derived from the observation of nonrotating NSs also provide an upper bound on the maximum spin of a NS. 
 The maximum spin frequency is given empirically as $f_{\rm lim} \simeq \frac{1}{2 \pi}  (0.468 + 0.378  \chi_{s}) \sqrt{\frac{G M_{\rm max}}{R_{\rm max}^3}}$,  
  \citep{1996ApJ...456..300L,Paschalidis:2016vmz} where $\chi_{s} = \frac{2 G M_{\rm max}}{R_{\rm max} c^2}$, with $M_{\rm max}$ and $R_{\rm max}$ being the  maximum mass and its corresponding radius of a nonrotating NS, respectively. 
 We use $M_{\rm max}-R_{\rm max}$ posterior samples that were deduced in~\citet{Biswas_arXiv_2008.01582B} by using PSR J0740+6620, combined GWs, and NICER data to calculate $f_{\rm lim}$. In the left panel of  Fig.~\ref{fig:rot-NS-prob}, its distribution is shown by the grey shaded region. We overlay that distribution with distributions of frequencies of the secondary object of GW190814 and those of a few hypothetical 
rotating NSs of various masses -- all Gaussian distributed, but with
medians of 2.4 $M_{\odot}$, 2.8 $M_{\odot}$ and 3.0 $M_{\odot}$, respectively, and each having a measurement uncertainty of $0.1 M_{\odot}$. We also assume the primary component of GW190425 to be a rapidly rotating NS, since by  using a high-spin prior LVC determined its mass to be $1.61 M_{\odot}-2.52 M_{\odot}$. In our calculations, for GW190425 we used the publicly available high-spin posterior of $m_1$ obtained by using the PhenomPNRT waveform. We find observations like $m_1$ of GW190425 and simulations like  $\mathcal{N} (2.4 M_{\odot},0.1 M_{\odot})$ correspond to posteriors of rotational frequency that are comparatively lower than limiting values of rotational frequencies. However, as the mass increases, the posterior of frequency eventually almost coincides with $f_{\rm lim}$. Therefore, if the secondary of GW190814 was a rapidly rotating NS, it  would have to be rotating rather close to the limiting frequency.

Any rotating star is generically unstable through the Chandrasekhar-Friedman-Schutz (CFS) mechanism~\citep{Chandrasekhar:1992pr,Friedman:1978hf}. This instability occurs when a certain retrograde mode in the rotating frame becomes prograde in the  inertial frame. For example, the $f$-modes of a rotating NS can always be made unstable for a sufficiently large mode number $m$ (not to be confused with component masses $m_{1,2}$) even  for low spin frequencies, but, the instability timescale increases rapidly with the increase of $m$.
Numerical calculations have shown~\citep{Stergioulas:1997ja,Morsink_1999}, that for maximum mass stars $m=2$ mode changes from retrograde to prograde at $T/|W| \sim 0.06$, where $T$ is the rotional energy and $W$ the gravitational potential energy of the NS. We computed this ratio for all the cases considered in this section and plot the distributions in the right panel of Fig.~\ref{fig:rot-NS-prob}. From this analysis we find that the secondary of GW190814 should be $f-$mode unstable as for most of the allowed EoSs $T/|W|$ is significantly larger than $0.06$.
The CFS instability is even more effective for $r-$modes~\citep{Lindblom:1998wf,1999ApJ...510..846A}
as they are generically unstable for all values of spin frequency. 
However, an instability can develop, only if its growth timescale is shorter than the timescale of the strongest damping mechanism affecting it. A multitude of damping mechanisms, such as shear viscosity, bulk viscosity, viscous boundary layer, crustal resonances and  superfluid mutual friction (each having each own temperature dependence) have been investigated (see \citep{2016EPJA...52...38K,Paschalidis:2016vmz,Andersson:2019yve,2021ApJ...910...62Z} and references therein). The spin-distribution of millisecond pulsars in accreting systems \citep{Papitto:2014yia} can be explained, if the $r$-mode instability is effectively damped up to  spin frequencies of $\sim 700\,$Hz  \citep{2011PhRvL.107j1101H},  and operating at higher spin rates. This would not allow for the secondary in GW190814 to be a rapidly rotating NS at the limiting spin frequency.  

On the other hand, if the secondary of GW190814 {\it was} a rapidly rotating NS at the limiting frequency, then the $f$-mode and $r$-mode instabilities must be effectively damped both during the spin-up phase in a low-mass-X-ray binary, where it acquires rapid rotation, as well as during its subsequent lifetime up to the moment of merger. This might be possible, if both the $f$-mode and the $r$-mode instabilities are damped by a particularly strong mutual friction of superfluid vortices below the superfluid transition temperature of $\sim 10^9$K (see \cite{ 2000PhRvD..61j4003L, 2011PhRvL.107j1102G} and in particular the case of an intermediate drag parameter ${\cal R}\sim 1$ in \citet{10.1111/j.1365-2966.2009.14963.x}). If this is the case, then the limiting frequency observed in the spin-distribution of millisecond pulsars must be explained by other mechanisms; see \citet{Gittins:2018cdw}. A possible presence of rapidly rotating NS in merging binaries thus would have strong implications on the physics of superfluidity  in NS matter (in particular constraining the drag parameter $\cal R$ of mutual friction) and on the astrophysics of accreting systems.

\section{Constraining NS EoS assuming that the GW190814 secondary is a BH.}
So far, we have analyzed the impact on NS EoS properties arising from the hypothesis that the secondary object in GW190814 is a NS. On the other hand, if that secondary object is a BH, then again novel information about the NS EoS can be obtained, since it will set an upper bound on the NS maximum mass, but only if one were to assume that the NS and BH mass distributions do not overlap.

\begin{figure}
    \centering
    \includegraphics[width=0.45\textwidth]{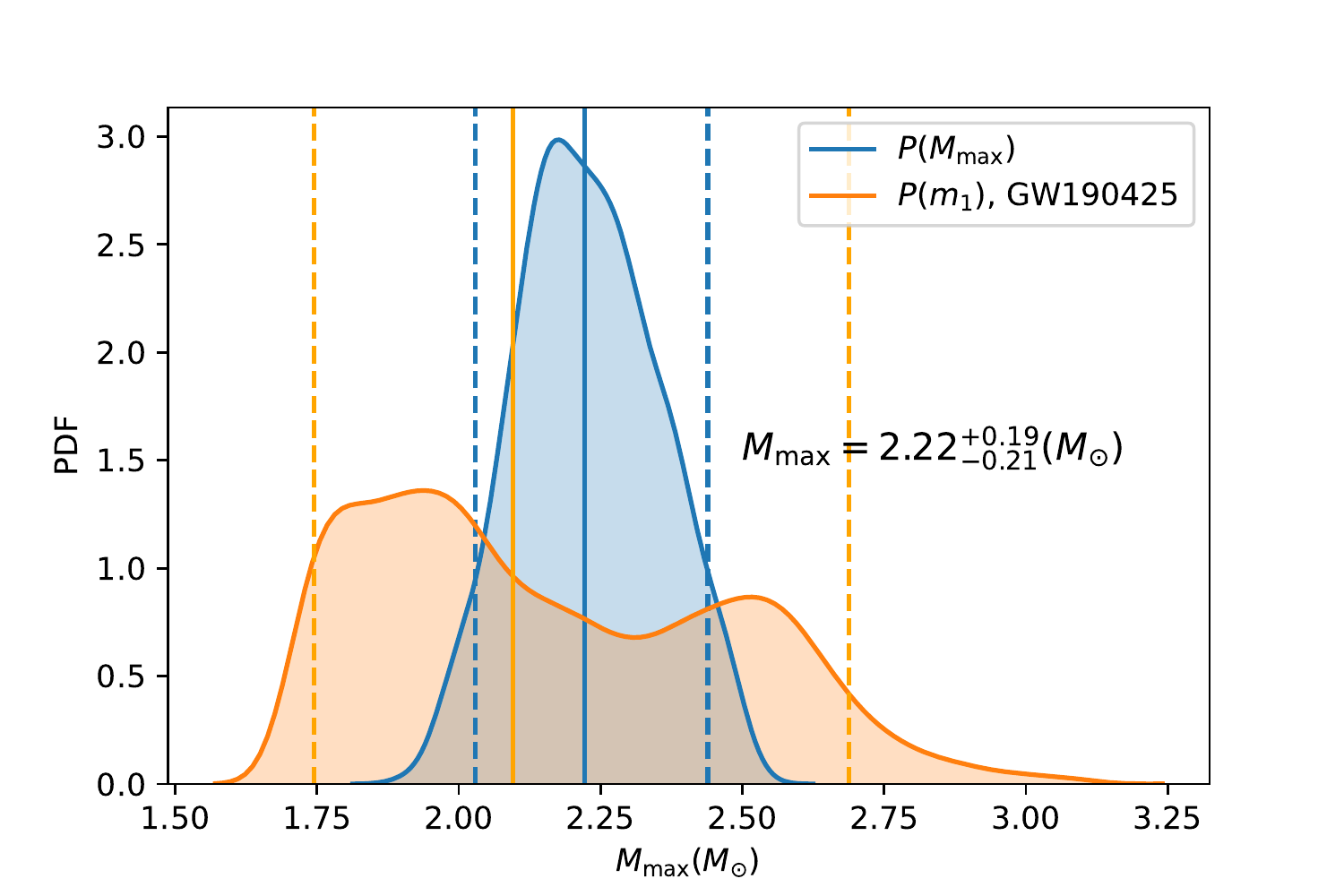}
\caption{The probability of NS $M_{\rm max}$ is plotted in blue,  under the hypothesis that the GW190814 secondary is a BH. Overlaid in orange is the LVC posterior of the primary in GW190425, for the high-spin prior.}

    \label{fig:mmax_GW190814_BBH}
\end{figure}
 In our analysis, we take this value to be $2.5 M_{\odot}$, which is the lowest possible value of the secondary object within $90 \%$ CI. Then, using Bayesian inference for nonrotating stars, we combine PSR J0740+6620, GWs, and NICER data to place further constraints on the NS EoS. In Fig.~\ref{fig:mmax_GW190814_BBH}, the distribution of the maximum mass for nonrotating NSs is shown in blue using the EoS samples obtained from this analysis. Within the $90 \%$ CI we find  $M_{\rm max}=2.22^{+0.19}_{-0.21} M_{\odot}$, which is the most conservative bound on NS maximum mass obtained so far in this work.

Assuming a high-spin prior, the mass of the primary component of GW190425 is constrained between $1.61-2.52 M_{\odot}$. In Fig.~\ref{fig:mmax_GW190814_BBH}, its distribution is over-plotted in orange. From the overlap with the newly obtained $M_{\rm max}$ distribution and the $m_1$ distribution of GW190425, we find that there is $\sim 40 \%$ probability that the primary of GW190425 is a BH. 

\section{Conclusion}
 Based on the maximum mass samples obtained from~\citet{Biswas_arXiv_2008.01582B}, we find that there is $\sim 1 \%$ probability that the secondary object associated with GW190814 is a nonrotating NS.
However, 
such an estimation
depends on the choice of EoS parameterization and the maximum mass threshold. Nevertheless, the possibility of the secondary being a nonrotating NS is not inconsistent with the data.
Based on our hybrid nuclear+PP EoS parameterization, we find that the addition of GW190814 as a nonrotating stars provides a very stringent constraint on the EoS specially in the high density region. We also discussed the alternative that the secondary is a rapidly rotating NS. We find that in order to satisfy the secondary mass estimate of GW190814, its spin magnitude has to be  close to the limiting spin frequency for uniform rotation. In fact, it would be the fastest rotating NS ever observed. However, this could be the case, only if gravitational-wave instabilities are effectively damped for rapidly rotating stars, which opens the possibility of constraining physical mechanisms, such as mutual friction in a superfluid interior.

\section*{Note added in proof} 
Recently the mass of PSR J0740+6620 was  revised slightly downwards -- to $2.08^{+0.07}_{-0.07} M_{\odot}$, at $68 \%$ CI~\citep{2021arXiv210400880F}. This has a marginal effect on the EoS posterior, potentially making some slightly softer EoSs viable~\citep{Biswas:2021yge}. Consequently, under the rapidly rotating scenario, the spin of the secondary would become slightly higher than what is reported in this paper.

\section*{Acknowledgements}
We  thank Philippe Landry and Toni Font for carefully reading the manuscript and making several useful suggestions. We gratefully acknowledge the use of high performance super-computing cluster Pegasus at IUCAA for this work.
P.C. is supported by the Fonds de la Recherche Scientifique-FNRS, Belgium, under grant No. 4.4503.19. This research has made use of data, software and/or web tools obtained from the Gravitational Wave Open Science Center (\href{https://www.gw-openscience.org}{https://www.gw-openscience.org}), a service of LIGO Laboratory, the LIGO Scientific Collaboration and the Virgo Collaboration. LIGO is funded by the U.S. National Science Foundation. 
The authors gratefully acknowledge the Italian Istituto Nazionale di Fisica Nucleare (INFN),  
the French Centre National de la Recherche Scientifique (CNRS) and
the Netherlands Organization for Scientific Research, 
for the construction and operation of the Virgo detector
and the creation and support  of the EGO consortium.
We would like to thank all of the essential workers who put their health at risk during the COVID-19 pandemic, without whom we would not have been able to complete this work.

\bibliographystyle{mnras}
\bibliography{mybiblio}
\label{lastpage}
\end{document}